\documentclass[graybox]{svmult}

\usepackage{type1cm}        
\usepackage{makeidx}         
\usepackage{graphicx}        
\usepackage{multicol}        
\usepackage[bottom]{footmisc}

\usepackage{newtxtext}       %
\usepackage[varvw]{newtxmath}       

\usepackage{bbm}

\makeindex             

\newcommand\ket[1]{\left|#1\right\rangle}

\newcommand{\tr}{\operatorname{Tr}}
\newcommand{\partr}{\operatorname{tr}_B}

\begin{document}

\title*{The role of initial coherence in the phase-space entropy production rate}
\author{Giorgio Zicari \and Bar{\i}\c{s} \c{C}akmak \and Mauro Paternostro}
\institute{Giorgio Zicari \at Centre for Quantum Materials and Technologies, School of Mathematics and Physics, Queen’s University, Belfast BT7 1NN, United Kingdom
\and Bar{\i}\c{s} \c{C}akmak  \at Department of Physics, Farmingdale State College - SUNY, Farmingdale, New York 11735, USA;
\newline
College of Engineering and Natural Sciences, Bahçeşehir University, Beşiktaş, Istanbul 34353, Turkey.
\and Mauro Paternostro  \at Centre for Quantum Materials and Technologies, School of Mathematics and Physics, Queen’s University, Belfast BT7 1NN, United Kingdom;
\newline
Universit\`{a} degli Studi di Palermo, Dipartimento di Fisica e Chimica – Emilio Segr\`{e}, via Archirafi 36, I-90123 Palermo, Italy
}
%
%
\maketitle

\abstract{The second law of thermodynamics can be expressed in terms of entropy production, which can be used to quantify the degree of irreversibility of a process. In this Chapter, we consider the  standard scenario of open quantum systems, where a system irreversibly interacts with an external environment. We show that the standard approach, based on von Neumann entropy, can be replaced by a phase-space formulation of the problem. In particular, we focus on spin systems that can be described using the so-called spin coherent states. We deploy this formalism to study the interplay between the entropy production rate and the initial quantum coherence available to the system.}

\section{Introduction}
\label{sec:intro}

First introduced by Maxwell in \emph{The Theory of Heat}, the well-known demon named after him is the protagonist of a famous though experiment that results in a paradoxical situation where one of the pillars of classical physics -- the second law of thermodynamics -- is apparently violated~\cite{Maxwell}. Such a  thought-provoking experiment has attracted a great deal of interest over the decades, especially given that the problem, using the language of information theory, has been rephrased in terms of information gain and feedback control~\cite{Landauer,Bennett}. The picture is further complicated if one moves into the quantum domain, where the backaction is intimately related to the quantum measurement problem.

However, without going into the technical details of the issue, one of the questions that needs to be addressed in the fist place is the following: \emph{which type of quantum entropy do we need to use?} 

When one studies the thermodynamics of quantum systems, the standard approach is to rely on von Neumann entropy, due to its relevance for information-theoretic purposes~\cite{Nielsen_chuang}. In this Chapter, we show that an alternative approach is available, borrowing some of the well-know tools of quantum optics: the quantum phase-space formalism can be used to describe the production of entropy within a given system and between the latter and the environment that surrounds it.

In classical thermodynamics, entropy production is indeed the key quantity that allows us to express the second law of thermodynamics, providing a way to characterise and quantify the irreversibility of thermodynamic processes~\cite{Landi:2021}. Let us consider the standard scenario where we deal with a macroscopic system surrounded by a thermal reservoir. Following the splitting that was put forward by Prigogine~\cite{Prigogine}, one can express the infinitesimal variation in entropy $d S$ as
\begin{equation}
\label{eq:prig1}
d S = d \Sigma - d \Phi \, ,
\end{equation}
where $d \Phi$ is the entropy that the system exchanges with its surroundings, while $d \Sigma$ is the entropy intrinsically produced by the processes taking place within the system. The second law of thermodynamics imposes a constraint on the sign of $d \Sigma$, which is always non-negative (i.e. $d \Sigma \ge 0$), where the equals sign holds for reversible processes, while $d \Sigma > 0$ for irreversible transformations. Since we are interested in non-equilibrium processes, from Eq.~(\ref{eq:prig1}) we can immediately obtain the following expression in terms of rates
\begin{equation}
\label{eq:prig2}
\frac{d S}{d t} = \dot{\Sigma}(t) - \dot{\Phi}(t) \, ,
\end{equation}
where $\dot{\Sigma}(t) \ge 0$ and $\dot{\Phi}(t)$ are the \emph{entropy production rate} and the \emph{entropy flux rate}, respectively. 

\section{Entropy production in open quantum systems}
\label{sec:EP}

\subsection{From closed to open systems: the emergence of irreversibility}
\label{sec:OQSs}

Let us consider a generic quantum system $S$ fully described by a time-independent Hamiltonian $H_S$. For such a system, any of its states - fully described by the density operator $\rho = \rho(t)$ - evolves over time according to the following dynamical equation
\begin{equation}
\label{eq:dyn_eqs}
\dot{\rho} = \mathcal{L} [\rho] \, , 
\end{equation}
where $\mathcal{L}$ is the Liouvillian (super)-operator acting on the the operator $\rho$, which we assume to have been prepared in a given initial state $\rho_0 = \rho(t=0)$.

In the ideal case where the system is completely isolated, Eq.~(\ref{eq:dyn_eqs}) assumes the form of a Liouville - von Neumann equation with 

\begin{equation}
\label{eq:Liouville-von Neumann}
\mathcal{L} [\rho] \equiv - i [H_S, \rho] \, .
\end{equation}

\begin{figure}[b]
\sidecaption
\includegraphics[scale=.5]{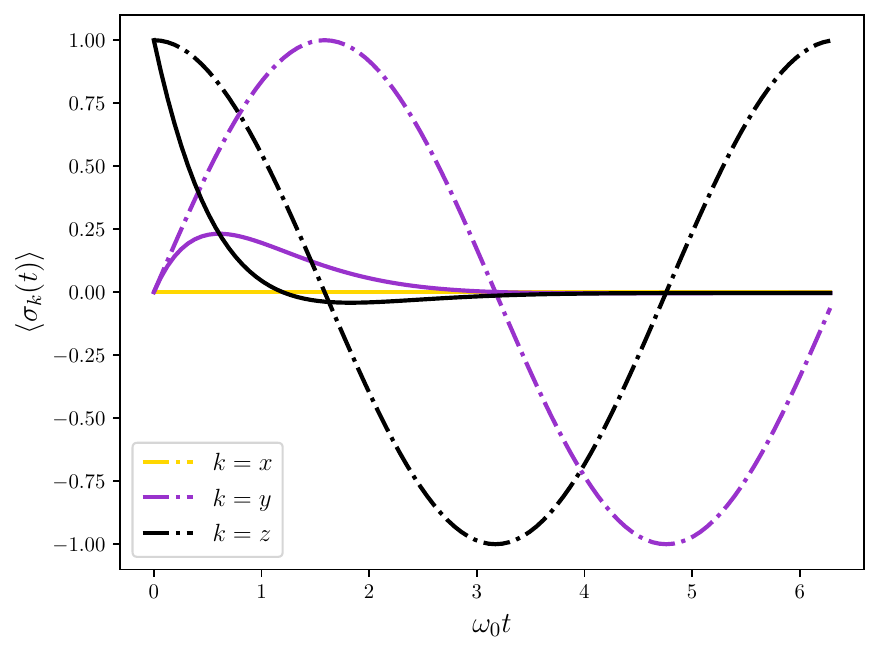}
\caption{Time evolution of the relevant observables  $\langle \sigma_k (t)\rangle$ (with $k = x,y,z$) of a qubit. The solid lines refer to the closed dynamics governed by Eq.~(\ref{eq:Liouville-von Neumann}), while dotted-dashed lines refer to the dissipative dynamics, as described by a dissipator of the form of Eq.~(\ref{eq:Davies-Lindblad_maps}). The plots clearly shows that the interaction with the thermal bath leads to irreversibility, causing the observables to depart from the periodic oscillations of the prefectly isolated scenario.}
\label{fig:qubit_dynamics}   
\end{figure}

 In this scenario, the system's dynamics is fully reversible: the equations of motion are invariant under the time-reversal symmetry transformation $t \to-t$. This can be seen straightforwardly by considering a two-level system (i.e. a qubit) with energy described by the Hamiltonian $H_S = (\omega_0/2) \sigma_x$\footnote{We  assume units such that $\hbar = k_B =1$.}. Here $\omega_0$ is the energy difference between the two levels, while $\sigma_j$ is the $j=x,y,z$ Pauli operator. The system, initially prepared in one of the eigenstates of $\sigma_z$, i.e. either $\ket{0}$ or $\ket{1}$, oscillates indefinitely between such states --- cf. Fig.~\ref{fig:qubit_dynamics}.

Now we let our system $S$ -- which we assume to be fully within our control -- interact with a finite (small) number $N$ of otherwise unaccessible quantum harmonic oscillators (hereafter dubbed {\it modes}), initially prepared in a thermal state at a given temperature $T$.
We are thus setting a boundary between $S$ and a second subsystem $B$, which can be called \textit{environment}, made out of the collection of modes. This has a precise physical meaning, as it implies focusing onto $S$ (the so-called \textit{reduced} system) while effectively discarding the environmental degrees of freedom. Mathematically, it is tantamount to performing a partial trace using  the basis of the Hilbert space associated with $B$. Therefore, by explicitly solving Eq.~(\ref{eq:Liouville-von Neumann}) for the state $\rho_{SB}(t)$ of the $S-B$ compound, we can obtain $\rho_S(t)= \partr\rho_{SB}(t)$ at any time $t$. By doing so, we depart from the totally reversible scenario we previously had when $S$ was perfectly isolated --- this becomes more and more evident as the number of modes $N$ increases. The truly irreversible dynamics is obtained when we take the limit $N\to \infty$.  

In principle, the operation of tracing out the environment can be done in a (numerically) exact way. However, this becomes quickly unfeasible as the number of modes increases. Luckily enough, the theory of open quantum systems provides us with an effective way to perform this task. By making some further assumptions -- collectively known as \textit{Born-Markov approximation}~\cite{Breuer-Petruccione} -- we are able to track the evolution of the reduced system. We should stress that this comes at a cost: we are coarse-graining the system dynamics assuming that the environmental correlation functions decay much faster than the typical timescale over which our system evolves. Physically, it means that the information is monotonically flowing from the system to the environment, leading to a Markovian (or memoryless) dynamical process~\cite{Breuer-Petruccione,Lindblad:1976}. Such a separation of the timescales allows us to derive special classes of dynamical equations, known as master equations, that can be still brought in the form of Eq.~(\ref{eq:dyn_eqs}), where the Liouvillian reads as 
\begin{equation}
\label{eq:Lindblad_ME}
\mathcal{L} [\rho] \equiv - i [H_S, \rho] + D(\rho) \, ,
\end{equation}
with the first term accounting for the unitary dynamics, while $D(\rho)$ being the so-called dissipator, which effectively includes the environmental effects, such as decoherence and/or dissipation. In particular, we would like to focus on dissipators of a special form, known as \textit{Lindblad-Davies maps}, which play a important role in quantum thermodynamics. They can be expressed as
\begin{equation}
\label{eq:Davies-Lindblad_maps}
D(\rho) = \sum_j \Gamma_j^- \left ( L_j^- \rho L_j^+ - \frac{1}{2} \{ L_j^+ L_j^-, \rho \} \right ) + \sum_j \Gamma_j^+ \left ( L_j^+ \rho L_j^- - \frac{1}{2} \{ L_j^- L_j^+, \rho \} \right ) \, ,
\end{equation}
where $L_j^\pm$ are the Lindblad operators describing the coupling between the system $S$ and the various decay channels with damping rates $\Gamma_j^\pm$~\cite{Breuer:2003}. The latter are assumed to satisfy the local detailed balance condition
\begin{equation}
\label{eq:detailed_balance}
\frac{\Gamma_j^+}{\Gamma_j^-} = e^{- \beta \omega_j}
\end{equation}
for each $j$, where $\beta$ is the inverse temperature of the thermal reservoir. Once again we can consider the example of a qubit, where we consider only two decay channels, with rates $\Gamma^- = \Gamma (\bar{n} +1)$ and $\Gamma^+ = \Gamma \bar{n}$, accounting for the incoherent loss and gain of excitations, respectively. Note that $\bar{n} = (e^{\beta \omega} - 1)^{-1}$ is the average number of excitations in the bath. The corresponding Lindblad operators are given by $L^{\pm} = \sigma_{\pm} = (\sigma_x \pm i \sigma_y)/2$. The time evolution of the observables clearly shows the irreversible character of the dynamics, as shown in Fig.~\ref{fig:qubit_dynamics}.

\subsection{Standard approach: von Neumann entropy production}

When we move from an isolated to an open system, we allow the main system to exchange energy -- or, generally speaking, \emph{information} - with its surroundings. To quantify the entropy production rate and flux, we need to introduce a suitable definition of entropy~\cite{Spohn:1977}. The standard choice for quantum information processes is given by the von Neumann entropy, which -- for a state $\rho$ -- is defined as 
\begin{equation}
S_{\rm vN}(\rho) = - \tr{ (\rho \ln \rho) } .
\end{equation}
By taking the time derivative of the latter, we readily obtain $\dot{S}_{\rm vN}(\rho) = - \tr{ (\dot{\rho} \ln \rho) }$. Since $\dot{\rho} = \mathcal{L} [\rho]$, we obtain
\begin{equation}
\label{eq:S_vn_dot}
\dot{S}_{\rm vN}(\rho) = - \tr{ \left ( \mathcal{L}[\rho] \ln \rho \right )} \, .
\end{equation}
The entropy flux rate is associated with the heat flow, which, in the weak coupling limit, is defined as $\dot{Q} \equiv - \tr (H_S \dot{\rho})$. If the system is weakly coupled to a thermal reservoir at inverse temperature $\beta$, the system relaxes towards the local Gibbs state given by $\rho_{\rm eq} = e^{-\beta H_S}/\mathcal{Z}_\beta$, where $\mathcal{Z}_\beta \equiv \tr ( {e^{-\beta H_S}})$ is the partition function.

As in the classical case, we have
\begin{equation}
\label{eq:phi_vn}
\dot{\Phi} (t) = \beta \, \dot{Q}(t) = \tr{ \left ( \mathcal{L}[\rho] \ln \rho_{\rm eq} \right )} \, ,
\end{equation}
where we have resorted to the identity $\ln \rho_{\rm eq} = - \beta H_S - \mathbbm{1} \ln \mathcal{Z}_\beta$ and the fact that $\tr{\dot{\rho}} = 0$.

Eqs.~(\ref{eq:S_vn_dot})-(\ref{eq:phi_vn}) lead to the following expression for the entropy production rate:
\begin{equation}
\label{eq:EP_vN1}
\dot{\Sigma}(t) = \dot{S}_{\rm vN}(\rho) + \dot{\Phi} (t) = -\tr{\left \{ \mathcal{L}[\rho] \left ( \ln \rho - \ln \rho_{\rm eq} \right )\right \}} \,
\end{equation}
which can be recast in the form of a time derivative, i.e.
\begin{equation}
\label{eq:EP_vN2}
\dot{\Sigma}(t) = - \frac{d}{d t} S(\rho || \rho_{\rm eq}) \, ,
\end{equation}
provided that we introduce the von Neumann relative entropy, defined as $S_{\rm vN}(\varrho || \varsigma) \equiv \tr{\{ \varrho (\ln \varrho - \ln \varsigma) \}}$. For a quantum system described by a Liouvillian in the form of Eq.~(\ref{eq:Lindblad_ME}), from the so-called Spohn inequality it follows that $\dot{\Sigma}(t) \ge 0$ at all times, as required by the second law of thermodynamics~\cite{Spohn:1977}. 
From Eqs.~(\ref{eq:phi_vn}) and (\ref{eq:EP_vN1}), it is immediate to conclude that $\dot{\Phi}(t) = 0 = \dot{\Sigma}(t)$ when the system reaches the local Gibbs state, as $\mathcal{L}[\rho_{\rm eq}] = 0$. 

\subsection{The role played by quantum coherence}
\label{sec:coherence}

It can be explicitly shown that a master equation of the form specified by Eq.~(\ref{eq:Lindblad_ME}) can be derived \emph{ab initio} starting from the microscopic Hamiltonian describing the system, its environment, and the interaction between them. By working in the Born-Markov approximation, one obtains two separate sets of dynamical equations: one governing the diagonal entries of $\rho$ (i.e. the \emph{populations}), the other governing the off-diagonal entries (i.e. the \emph{coherences})~\cite{Breuer-Petruccione}. This mathematical evidence expresses the occurrence of two different physical processes dynamically taking place: on one hand, there are transitions between the energy levels, causing the populations to adjust to values imposed by the thermal bath; on the other hand, one witnesses the loss of coherence in the energy eigenbasis.

For the sake of definiteness, let us consider the eigenbasis $\{ \ket{n} \}$. To be more quantitative, the populations $P_n \equiv \langle n| \rho | n \rangle $ behave classically, obeying the Pauli master equation~\cite{Gardiner:2009}
\begin{equation}
\label{eq:Pauli_ME}
\frac{d P_n}{dt} = \sum_n \left [ W(n|k) P_k(t) - W(k|n) P_n(t)  \right] \, ,
\end{equation}
where $W(n|k)$ are the transition rates from the energy level $\epsilon_n$ to the level $\epsilon_k$, satisfying the detailed balance condition
\begin{equation}
\label{eq:detailed_balance}
\frac{W(n|k)}{W(k|n)} = \frac{P_n^{\rm eq}}{P_k^{\rm eq}} = e^{- \beta (\epsilon_n - \epsilon_k)} \, ,
\end{equation}
where $P_n^{\rm eq} \equiv \langle n| \rho_{\rm eq} | n \rangle$.

Inspired by the general definition of free energy, i.e. $F = E - T S $, we can define the following non-equilibrium free energy~\cite{Santos:2019}
\begin{equation}
F(\rho) = \tr{\left ( H_S \rho\right )} + T \tr{\left (\rho \ln \rho \right)} \, 
\end{equation}
where $T$ is the temperature of the bath. It is immediate to check that, for equilibrium states, one recovers the well known result of statistical mechanics
\begin{equation}
 F_{\rm eq} \equiv F(\rho_{\rm eq}) = -T \ln \mathcal{Z}_{\beta} \, ,
\end{equation}
whose generalisation for non-equilibrium states reads as
\begin{equation}
\label{eq:non_eq_F}
F(\rho) = F_{\rm eq} + T S(\rho || \rho_{\rm eq}) \, .
\end{equation}
Since $S(\rho || \rho_{\rm eq}) \ge 0$, we readily obtain the condition $F(\rho) \ge F_{\rm eq}$, which automatically defines the equilibrium state $\rho_{\rm  eq}$ as the one minimising the free energy. 

Combining Eqs.~(\ref{eq:EP_vN2}) and (\ref{eq:non_eq_F}), we obtain the following expression for the entropy production rate
\begin{equation}
\dot{\Sigma}(t) = - \frac{1}{T} \frac{d F(\rho)}{d t} \, .
\end{equation}

Note that the state $\rho_{\rm eq}$ is diagonal in the eigenbasis $\{ \ket{n} \}$ of $H_S$; differently, the state $\rho$ is in general an incoherent state which can be decomposed into a diagonal part $\rho_{\rm diag}$ and a off-diagonal one $\chi$, i.e. $\rho = \rho_{\rm diag} + \chi$. Therefore, we can introduce the relative entropy of coherence~\cite{Baumgratz:2014}, defined as
\begin{align}
C(\rho) = S(\rho_{\rm diag}) - S(\rho) \, , 
\end{align}
which allows us to split the relative entropy as
\begin{equation}
S(\rho || \rho_{\rm eq}) = S(\rho_{\rm diag} || \rho_{\rm eq}) + C(\rho) \, ,
\end{equation}
where $S(\rho_{\rm diag} || \rho_{\rm eq})$ is the Kullback-Leibler divergence of the classical probability distribution $\{ P_n \}_n$ relative to the equilibrium distribution $\{ P_n^{\rm eq} \}_n$, i.e.
\begin{equation}
    \label{eq:KL}
    S(\rho_{\rm diag} || \rho_{\rm eq}) = \sum_n P_n \ln \left ( \frac{P_n}{P_n^{\rm eq}}\right) \, .
\end{equation}
Therefore, Eq.~(\ref{eq:non_eq_F}) can be rewritten as 
\begin{equation}
\label{eq:non_eq_F_2}
F(\rho) = F_{\rm eq} + T S(\rho_{\rm diag} || \rho_{\rm eq}) + T C(\rho) \, ,
\end{equation}
which shows that, apart from the equilibrium contribution $F_{\rm eq}$, the non-equilibrium free energy is made of two separate contributions: one classical and one genuinely quantum, corresponding to the second and third term of Eq.~(\ref{eq:non_eq_F_2}), respectively. The former quantifies the increase in free energy due to the population imbalance with respect to the equilibrium configuration; the latter, instead, expresses the additional free energy contribution carried by a state with non-null coherences~\cite{Santos:2019}.

It is straightforward to show that the splitting introduced by Eq.~(\ref{eq:non_eq_F_2}) carries over to the entropy production rate
\begin{equation}
\dot{\Sigma}(t) = -\frac{d}{dt} S(\rho_{\rm diag} || \rho_{\rm eq}) - \frac{d C(\rho)}{dt} \equiv \dot{\Sigma}_d(t) + \Upsilon(t) \, ,
\end{equation}
where, by resorting to Eq.~(\ref{eq:KL}), we obtain the following expression for the classical contribution to the entropy production rate
\begin{equation}
   \dot{\Sigma}_d(t) = \frac{1}{2} \sum_{n,k} \left [ W(n|k) P_k -  W(k|n) P_n \right ] \ln \left ( \frac{W(k|n) P_n}{W(n|k) P_k} \right ) \, ,
\end{equation}
where we have used the Pauli master equation~(\ref{eq:Pauli_ME}) together with the detailed balance condition~(\ref{eq:detailed_balance}). The contribution due to quantum coherence is simply given by $\Upsilon(t) = - d C(t)/ dt$.

\section{Entropy production in the quantum phase space}

The standard approach to entropy production in open quantum systems described in Sect.~\ref{sec:EP} provides a consistent framework based on von Neumann entropy. We should stress that the aforementioned results are based on a certain number of identifications and definitions which are valid if and only if we deal with quantum systems weakly coupled with a thermal reservoir, as far as we satisfy all the assumptions of the Born-Markov approximation. However, even when we move within that approximation, the framework we have just introduced is not flawless. For instance, in the limit of zero temperature reservoirs (i.e. $\beta \to \infty$) this framework is formally inconsistent. On one hand, the reference state in the relative entropy, namely $\rho_{\rm eq}$, becomes pure, which causes $S(\rho || \rho_{\rm eq})$ and $\dot{\Sigma}(t)$ [cf. Eq.~(\ref{eq:EP_vN2})] to be ill-defined. The entropy flux rate $\dot{\Phi}(t)$ is also divergent, as one can easily conclude from Eq.~(\ref{eq:phi_vn}). However, the overall rate $d S / dt $ stays finite, showing that the aforementioned divergences have no corresponding physical meaning: the zero temperature limit is indeed frequently considered in quantum optics~\cite{Brunelli:2018}. A similar inconsistency is found whenever the system is in a pure state.

Therefore, one possible solution to this problem rely on a semi-classical formulation for the entropy production in the quantum phase-space, which coincides with the standard approach at high temperatures. The core idea is to replace the von Neumann entropy with a generalised entropy function, associated with a certain probability distribution defined over the phase space~\cite{Santos:2017,Santos:2018}. In particular, hereafter we will refer to the Wehrl entropy defined using the Husimi-Q function~\cite{Wehrl}. We will resort to the latter to describe spin systems for which a phase-space formulation is available, namely those described through the so-called spin coherent states.

\subsection{Spin coherent states}
\label{sec:spin_coherent_states}

Let us consider a single quantum system described by the operators $J_x, J_y, J_z$ satistying the algebra $[J_x, J_y] = i J_z$. A \emph{spin coherent state} is defined as~\cite{Spin_Coherent_States}
\begin{equation}
\label{eq:spin_coherent_state_def}
\ket{\Omega} = e^{- i \phi J_z} e^{- i \theta J_y} \ket{J} \, ,
\end{equation}
where $\Omega = (\theta, \phi)$ is a solid angle in polar coordinates (with $\theta \in [0,\pi]$ and $\phi \in [0, 2 \pi]$), while $\ket{J}$ is the angular momentum state with the largest quantum number of $J_z$.

If the system is described by the density operator $\rho$, we can define the Husimi Q-function $\mathcal{Q} = \mathcal{Q}(\Omega)$ over the phase space
\begin{equation}
\label{eq:HusimiQ}
\mathcal{Q}(\Omega) \equiv  \langle \Omega | \rho | \Omega \rangle.
\end{equation}
Furthermore, if the system dynamics is governed by a master equation in the Lindblad form, as in Eq.~(\ref{eq:Lindblad_ME}), the phase-space counterpart of latter assumes the form of a Fokker-Planck equation, i.e.
\begin{equation}
\label{eq:FP_eq}
\partial_t \mathcal{Q} = \mathcal{U} (\mathcal{Q}) + \mathcal{D} (\mathcal{Q}) \, ,
\end{equation}
where $\mathcal{U} (\mathcal{Q})$ and $\mathcal{D} (\mathcal{Q})$ account for the unitary and dissipative parts of the evolution, respectively.

If the system evolves according to a standard open system dynamics described by the generic Liouvillian in Eq.~(\ref{eq:Lindblad_ME}), one can derive a Fokker-Planck equation of the form~(\ref{eq:FP_eq}) using a set of suitable correspondence rules. For instance, we have~\cite{Spin_Coherent_States}
\begin{align}
\label{eq:correspondence_rules}
\left[J_{+} , \rho \right ] &\rightarrow \mathcal{J}_{+} (\mathcal{Q}) = e^{i \phi} \left (\partial_{\theta} + i \cot{\theta} \; \partial_{\phi} \right ) \mathcal{Q}\, , \nonumber \\
\left[J_{-} , \rho \right ] & \rightarrow \mathcal{J}_{-} (\mathcal{Q}) = -e^{-i \phi} \left (\partial_{\theta} - i \cot{\theta} \; \partial_{\phi} \right )\mathcal{Q}\, , \nonumber\\
\left[J_{z} , \rho \right ] & \rightarrow \mathcal{J}_{z} (\mathcal{Q}) = - i {\partial}_{\phi} \mathcal{Q}\, ,
\end{align}
where $J_{\pm} = J_x \pm i J_y$.

\subsection{Spin-phase-space entropy production}

Given the Husimi Q-function $\mathcal{Q} = \mathcal{Q}(\Omega)$, one can define the Wehrl entropy as~\cite{Wehrl}
\begin{equation}
\label{eq:Wehrl_entropy}
\mathcal{S}_{\mathcal{Q}} = - \left ( \frac{2J + 1}{4 \pi} \right ) \int d \Omega \, \mathcal{Q} \ln \mathcal{Q} \, ,
\end{equation}
where the prefactor is chosen just for convenience. By taking the time derivative of Eq.~(\ref{eq:Wehrl_entropy}), the normalisation condition together with Eq.~(\ref{eq:FP_eq}) yields
\begin{equation}
\label{eq:Wehrl_entropy_rate}
\frac{d \mathcal{S}_{\mathcal{Q}}}{d t} \bigg |_{\rm diss} = -  \left ( \frac{2J + 1}{4 \pi} \right ) \int d \Omega \, \mathcal{D}(\mathcal{Q}) \ln \mathcal{Q} \, ,
\end{equation}
where we consider only the contribution coming from the dissipative part~\cite{Santos:2018}. For a given dynamical process, the idea is to  bring Eq.~(\ref{eq:Wehrl_entropy_rate}) into the Prigogine form of Eq.~(\ref{eq:prig2}), separating the entropy production rate $\dot{\Sigma}(t)$ from the flux rate $\dot{\Phi}(t)$.

\section{Spin-phase-space entropy production rate}
\label{sec:EP_phase_space}

In this Section, we derive the explicit expressions for the entropy production rate for two relevant classes of open system dynamics: dephasing and amplitude damping channels~\cite{Nielsen_chuang}. In order to do that, we start from Eq.~(\ref{eq:Wehrl_entropy_rate}) where we include only the contribution coming from the dissipative part of the dynamics, then we identify $\dot{\Sigma}$ and $\dot{\Phi}$ according to the splitting introduced in Eq.~(\ref{eq:prig2}).

\subsection{Dephasing channels}
\label{sec:deph}

Let us consider the case of dephasing channels, where the dissipator reads as
\begin{equation}
\label{eq:diss_deph}
D(\rho) = - \frac{\lambda}{2} \left [ J_z, \left [ J_z, \rho \right ] \right] \, ,
\end{equation}
with $\lambda$ being the dephasing rate. Note that a dissipator of this form does not entail any direct energy exchange between the system and the bath: the interaction between them causes the system to loose coherence. Using the correspondence rules (\ref{eq:correspondence_rules}), one can obtain the following phase-space dissipator
\begin{equation}
    \label{eq:diss_deph_phasespace}
    \mathcal{D}(\mathcal{Q}) = - \frac{\lambda}{2} \mathcal{J}_z \left ( \mathcal{J}_z (\mathcal{Q}) \right ) \, .
\end{equation}
By plugging Eq.~(\ref{eq:diss_deph_phasespace}) into Eq.~(\ref{eq:Wehrl_entropy_rate}), we get
\begin{equation}
\label{eq:EP_deph}
\dot{\Sigma} \equiv \frac{d\mathcal{S}_{\mathcal{Q}}}{dt} \biggl |_{\rm diss} = \frac{\lambda}{2} \left ( \frac{2 J +1}{4 \pi}\right ) \int \frac{|\mathcal{J}_z(\mathcal{Q})|^2}{\mathcal{Q}} d \Omega  \, ,
\end{equation}
after having performed an integration by parts. From Eq.~(\ref{eq:EP_deph}), it is immediate to conclude that $\dot{\Sigma} \ge 0$, as requested from the second law of thermodynamics. Note that $\dot{\Sigma} = 0$ if and only if $\mathcal{J}_z (\mathcal{Q}) = 0$, which occurs when the function $\mathcal{Q}=\mathcal{Q}(\theta,\phi)$ is independent of the azimuthal angle $\phi$, as, by definition, we have $\mathcal{J}_z (\mathcal{Q}) = - i \partial_\phi \mathcal{Q}$. In the phase space, the latter is equivalent to the condition of $\rho$ being diagonal in the $J_z$ basis. Therefore, $\mathcal{J}_z(\mathcal{Q})$ is the phase-space current associated to the genuine quantum phenomenon of decoherence. Moreover, it should be stressed that, for this channel, the identification of  $\dot{\Sigma}$ is straightforward, as the entropy flux rate $\dot{\Phi}$ is identically zero, due to the lack of energy exchange between the system and the bath.

\subsection{Amplitude damping channels}
\label{sec:ampl_damp}

We consider a second relevant example, i.e. amplitude damping channels, where the physical picture is richer: during the dynamical process, the populations adjust to values imposed by the thermal bath, while we witness the incoherent exchange of thermal excitations between the system and the environment. Such a process is mathematically expressed by a dissipator of the form  

\begin{equation}
\label{eq:ampl_damp_diss}
D(\rho) =  \Gamma (\bar{n} + 1)\left ( J_{-} \rho J_+ - \frac{1}{2} \{ J_{+} J_{-}, \rho \} \right ) + \Gamma \bar{n} \left ( J_{+} \rho J_{-} - \frac{1}{2} \{ J_{-} J_{+}, \rho \} \right ) \, ,
\end{equation}
where $\Gamma$ is the damping rate, while $\bar{n} = (e^{\beta \omega_0}-1)^{-1}$ is the average number of thermal excitations in the bath. Note that this dissipator is a prototypical example of a Lindblad-Davis map, as in Eq.~(\ref{eq:Davies-Lindblad_maps}). A more involved derivation (which can be found in Ref.~\cite{Santos:2018}) shows that the phase-space dissipator reads as
\begin{equation}
\label{eq:ampl_damp_diss_phase_space}
\mathcal{D}(\mathcal{Q}) = \frac{\Gamma}{2} \left \{ \mathcal{J}_{-}(f(\mathcal{Q})) - \mathcal{J}_{+}(f^*(\mathcal{Q}))\right \} \, ,
\end{equation}
where the phase-phase operators $\mathcal{J}_{\pm}$ are defined in Eq.~(\ref{eq:correspondence_rules}), while
\begin{equation}
f(\mathcal{Q}) = \frac{1}{2} \left [ 2 J \mathcal{Q} - \mathcal{J}_{z} (\mathcal{Q})\right ] e^{i \phi} \sin{\theta} + \frac{1}{2} \left [ \cos{\theta} - (2 \bar{n} + 1)\right] \mathcal{J}_{+}(\mathcal{Q}) \ .
\end{equation}
By substituting Eq.~(\ref{eq:ampl_damp_diss_phase_space}) into Eq.~(\ref{eq:Wehrl_entropy_rate}) and integrating by parts, we obtain
\begin{equation}
\frac{d \mathcal{S}_{\mathcal{Q}}}{dt} \biggl |_{\rm diss} = \frac{\Gamma}{2} \left ( \frac{2J +1 }{4 \pi}\right) \int \frac{\mathcal{F}(\mathcal{Q})}{\mathcal{Q}} d \Omega \, ,
\end{equation}
where $\mathcal{F}(\mathcal{Q}) = f(\mathcal{Q}) \mathcal{J}_{-}(\mathcal{Q}) - f^*(\mathcal{Q})\mathcal{J}_{+}(\mathcal{Q})$.

We can then separate $\dot{\Sigma}$ from $\dot{\Phi}$ as in Eq.~(\ref{eq:prig2}). The rationale behind this choice is that the entropy production and flux rates should be even and odd functions of the relevant currents, respectively, as expected from standard non-equilibrium thermodynamics arguments~\cite{Prigogine}. The calculations, detailed in Ref.~\cite{Santos:2018}, lead to the following expression for the entropy production rate
\begin{align}
\label{eq:EP_ampl_damp}
\dot{\Sigma} = \frac{\Gamma}{2} \left ( \frac{2J +1}{4 \pi}\right ) \int \frac{d \Omega}{ \mathcal{Q}} \biggl \{ \frac{\{ 2J \mathcal{Q} \sin{\theta} + [\cos{\theta} - (2 \bar{n} + 1)] \partial_{\theta} \mathcal{Q}\}^2}{(2 \bar{n} + 1) - \cos{\theta}} \nonumber \\
 + \left | \mathcal{J}_z(\mathcal{Q}) \right |^2 \left [  (2 \bar{n} +1) \cos{\theta} - 1\right] \frac{\cos{\theta}}{\sin^2{\theta}} \biggl \} \, .
\end{align}
This rather cumbersome expression for the entropy production rate is process-specific, in the sense that it applies to the specific type of dynamics we have considered. As a consequence, the mathematical expression contained in Eq.~(\ref{eq:EP_ampl_damp}) provides some insight on the processes taking place during the time evolution. It is indeed clear that there are two distinct contributions: one is proportional to the dephasing current $|\mathcal{J}_z(\mathcal{Q})|^2$, which accounts for the loss of coherence; the other directly related to the amplitude damping.

\section{Influence of the initial coherence on the entropy production rate}
\label{sec:EP_initial_coherence}

In this Section, we aim to study the influence of the initial coherence on the spin-phase-space entropy production rate. To this end, we will consider the expressions for the spin-phase entropy production rate derived in Sect.~\ref{sec:EP_phase_space}. For a given dynamical process (either dephasing or amplitude damping channels), we initially prepare the system in a state $\rho_0 = \rho_{\rm diag} + \chi$, where we explicitly separate the populations $\rho_{\rm diag}$ (i.e. the classical part) from the coherences $\chi$ (i.e. the quantum part). To quantify the initial coherence available to the system we use the $l_1$-norm~\cite{Baumgratz:2014}, i.e.
\begin{align}
\label{eq:coherence}
\mathcal{C}(\rho_0) = \sum_{i,j} |\chi_{ij}| \, .
\end{align}

We then solve the dynamical equations $\dot{\rho} = \mathcal{L}[\rho]$, whence we obtain the time evolution of the entropy production rate, i.e., $\dot{\Sigma} = \dot{\Sigma}(t)$.

\subsection{Qubit}
\label{eq:deph_init_coh}

Let us consider the simplest case of a single qubit (i.e. a single spin$-1/2$ particle), for which the most general density matrix can be written in the form
\begin{equation}
\label{eq:rho_Bloch}
\rho = \frac{1}{2} \left ( \mathbbm{1} + \boldsymbol{\sigma} \cdot \boldsymbol{\tau} \right ) \, ,
\end{equation}
where $\boldsymbol{\sigma} = (\sigma_x, \sigma_y, \sigma_z)$ and $\boldsymbol{\tau} = (\tau_x, \tau_y, \tau_z)$ are the Pauli and Bloch vectors, respectively, where $\tau_i = \tr(\sigma_i \rho)$. We can construct the spin coherent state $\ket{\Omega}$ using the definition in Eq.~(\ref{eq:spin_coherent_state_def}), where we take $J_i = \sigma_i/2$, and $\ket{J} = \ket{0}$ is the eigenstate of $\sigma_z$ corresponding to the eigenvector $+1$. From Eq.~(\ref{eq:HusimiQ}), the Husimi Q-function reads as
\begin{equation}
\label{eq:spin12_Husimi}
\mathcal{Q}(\Omega) = \frac{1}{2} \left ( \mathbbm{1} + \hat{n} \cdot \boldsymbol{\tau} \right ) \, ,
\end{equation}
where $\hat{n} = (\sin{\theta}\cos{\phi}, \sin{\theta} \sin{\phi}, \cos{\theta})$ is the unit vector.
Using Eq.~(\ref{eq:coherence}), the coherence available to the initial state can be immediately quantified as $\mathcal{C}(\rho_0) = 2(\tau_x^2 + \tau_y^2)$, where $\tau_x$ and $\tau_y$ are the first two Bloch components of the density matrix $\rho_0$ at time $t=0$. 

Let us consider the case of purely dephasing dynamics expressed by the dissipator in Eq.~(\ref{eq:diss_deph}), where $J_z = \sigma_z/2$. By plugging Eq.~(\ref{eq:spin12_Husimi}) into Eq.~(\ref{eq:EP_deph}), a cumbersome integration over the phase-space yields~\cite{Santos:2018}
\begin{equation}
\label{eq:EP_deph_spin12}
\dot{\Sigma} = \frac{\lambda}{4} (\tau_x^2 + \tau_y^2) \left \{ \frac{\tau - (1 - \tau^2)\tanh^{-1}(\tau)}{\tau^3} \right \} \, ,
\end{equation}
where $\tau = (\tau_x^2 + \tau_y^2 + \tau_z^2)^{1/2}$. By contrast, the von Neumann entropy production is given by~\cite{Santos:2018}
\begin{equation}
\label{eq:EP_deph_spin12_vN}
\dot{\Sigma}_{\rm vN} = \frac{\lambda}{2} (\tau_x^2 + \tau_y^2) \frac{\tanh^{-1}(\tau)}{\tau} \, .
\end{equation}
Note that, if the state of the system is pure, i.e. $\tau \to 1$, while the von Neumann entropy production diverges, the phase-space counterpart stays finite, namely $\dot{\Sigma} \to \lambda \sin^2 \theta / 4$, consistently with what stated at the beginning of Sect.~\ref{sec:EP_phase_space}. 

Analogously, one can derive the closed expression of $\dot{\Sigma}$ for the amplitude damping channels, by considering Eq.~(\ref{eq:ampl_damp_diss}) with $J_{\pm} = \sigma_{\pm}$. By plugging Eq.~(\ref{eq:spin12_Husimi}) into Eq.~(\ref{eq:EP_ampl_damp}), after integration, we eventually get~\cite{Santos:2018}
\begin{align}
\label{eq:EP_ampl_damp_spin12}
\dot{\Sigma} = \frac{\Gamma}{2} \frac{2 \bar{\tau}_z \tau_z - (\tau^2 + \tau_z^2)}{2 \bar{\tau}_z} & \left [ \frac{\tau - (1 - \tau^2) \tanh^{-1}(\tau)}{\tau^3} \right ] \nonumber \\
& + \frac{\Gamma}{2} (\tau_z - \bar{\tau}_z) \left [ \frac{\bar{\tau}_z - (1 - \bar{\tau}^2_z) \tanh^{-1}(\bar{\tau}_z)}{\bar{\tau}_z^3}
\right] \, ,
\end{align}
where we have introduced the bath-induced magnetisation $\bar{\tau}_z \equiv -1/(2 \bar{n}+1)$. Differently, one can show that the von Neumann entropy production rate reads as~\cite{Santos:2018}
\begin{equation}
\label{eq:EP_ampl_damp_spin12_vN}
\dot{\Sigma}_{\rm vN} = \Gamma \frac{\tanh^{-1}(\bar{\tau}_z)}{\bar{\tau}_z} (\tau_z - \bar{\tau}_z) - \frac{\Gamma}{2} \frac{\tanh^{-1}(\tau)}{\tau \bar{\tau}_z} \left [ \tau^2 + \tau_z (\tau_z - 2 \bar{\tau}_z) \right].
\end{equation}

\begin{figure}[b]
\includegraphics[scale=.45]{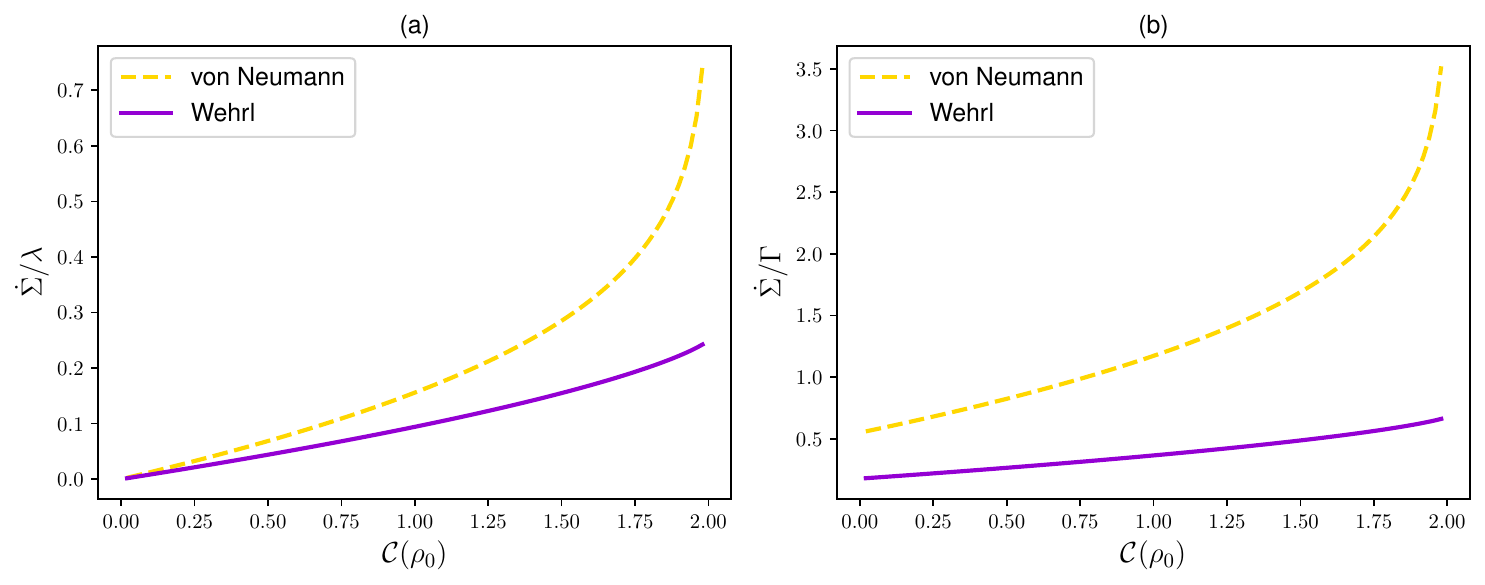}
\caption{Entropy production rate $\dot{\Sigma}$ as a function of the coherence available to the system, as measured through the $l_1$-norm quantifier, i.e. $\mathcal{C}(\rho_0) = 2(\tau_x^2 + \tau_y^2)$. In Panel (a), we consider the case of a system undergoing pure dephasing dynamics, as described by the dissipator in Eq.~(\ref{eq:diss_deph}). In Panel (b), we consider the case of amplitude damping channels described by Eq.~(\ref{eq:ampl_damp_diss}) instead. In both cases, we compare the von Neumann entropy production rate and its Wehrl counterpart. Note that the latter underestimates the entropy production rate, while the former diverges whenever we approach the limit of a pure state. For the the plots displayed in figure, we chose $\tau_z=0$ and $\bar{\tau}_z=0$.}
\label{fig:qubit_EP_vs_C}   
\end{figure}

For both classes of channels, we can plot the entropy production rates $\dot{\Sigma}$ and $\dot{\Sigma}_{\rm vN}$ as functions of the initial coherence $\mathcal{C}(\rho_0)$.  Fig.~\ref{fig:qubit_EP_vs_C} highlights the monotonic relationship between the quantum coherence we input and the entropy production rate.

\begin{figure}[b]
\includegraphics[scale=.45]{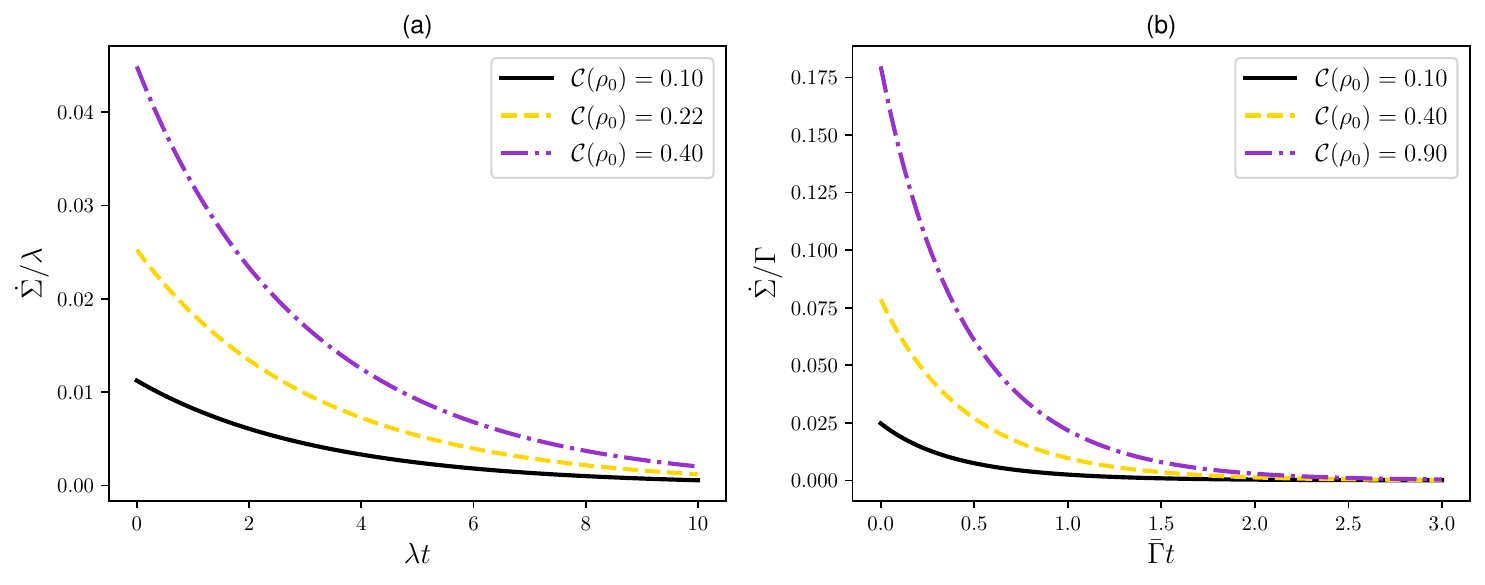}
\caption{Wehrl entropy production rate $\dot{\Sigma}$ as a function of time. In both Panels, we show several curves corresponding to different values of the initial coherence, as quantified by the $l_1$-norm, i.e. $\mathcal{C}(\rho_0) = 2(\tau_x^2 + \tau_y^2)$. In Panel (a), we consider a single qubit subject to pure dephasing, while in Panel (b) we consider we the case of amplitude damping dynamics. For the plots shown in Panel (a), we took different values of $\tau_x$ and $\tau_y$ to reproduce the values of coherence displayed in the legend, while $\tau_z$ has been taken so that $\tau^2 = 0.9$. In Panel (b), we chose $\tau_x$ and $\tau_y$ with the same rationale, while $\tau_z=0.1$. The average number of excitations in the bath is given by $\bar{n}=0.5$, while $\bar{\Gamma}=\Gamma (2\bar{n}+1)$.}
\label{fig:qubit_EP_vs_t}   
\end{figure}

One can also look at the dynamics of the entropy production rate. For the dephasing channel, the explicit solution of the dynamics in the interaction picture yields
\begin{equation}
\tau_i (t) = \tau_i(0) e^{- \lambda t/2} \quad i=x,y \, ,
\end{equation}
while the third component stays constant over time, i.e. $\tau_z(t) = \tau_z(0)$. By plugging $\boldsymbol{\tau}(t) = (\tau_x(t),\tau_y(t),\tau_z(t))$ back into Eqs.~(\ref{eq:EP_deph_spin12}), one can track the evolution of the phase-space entropy production rate. In Fig.~\ref{fig:qubit_EP_vs_t}-(a), we prepare the system in different initial states, labelled by a diffent value of the intial coherence $\mathcal{C}(\rho_0)$: one can immediately conclude that the higher is the initial coherence, the higher the entropy production rate.

Similarly, one can solve the amplitude damping dynamics, where all the components of the Bloch vector evolve non-trivially to eventually reach thermalisation. Explicitly, one gets
\begin{align}
    \label{eq:Bloch_vector_dyn_ampl_damp}
    \tau_i(t) & = \tau_i(0) e^{-\frac{\Gamma}{2} (2 \bar{n} +1) t} \, \quad i = x,y \, , \\
    \tau_z(t) & = \tau_z(0) e^{- \Gamma (2 \bar{n}+1) t } +\bar{\tau}_z \, .
\end{align}
If we plug these equations into Eqs.~(\ref{eq:EP_ampl_damp_spin12}), we obtain $\dot{\Sigma} = \dot{\Sigma}(t)$. Similarly to the case of pure dephasing channels, we consider different initial states, namely, different $\tau_x$ and $\tau_y$, while we keep $\tau_z$ fixed, as the latter enters in the diagonal part of the density matrix, which behaves classically. This case shows a similar behaviour: there is a monotonicity relationship between the coherence we input and the entropy production rate. 

\subsection{Qutrit}

\begin{figure}[b]
\includegraphics[scale=.45]{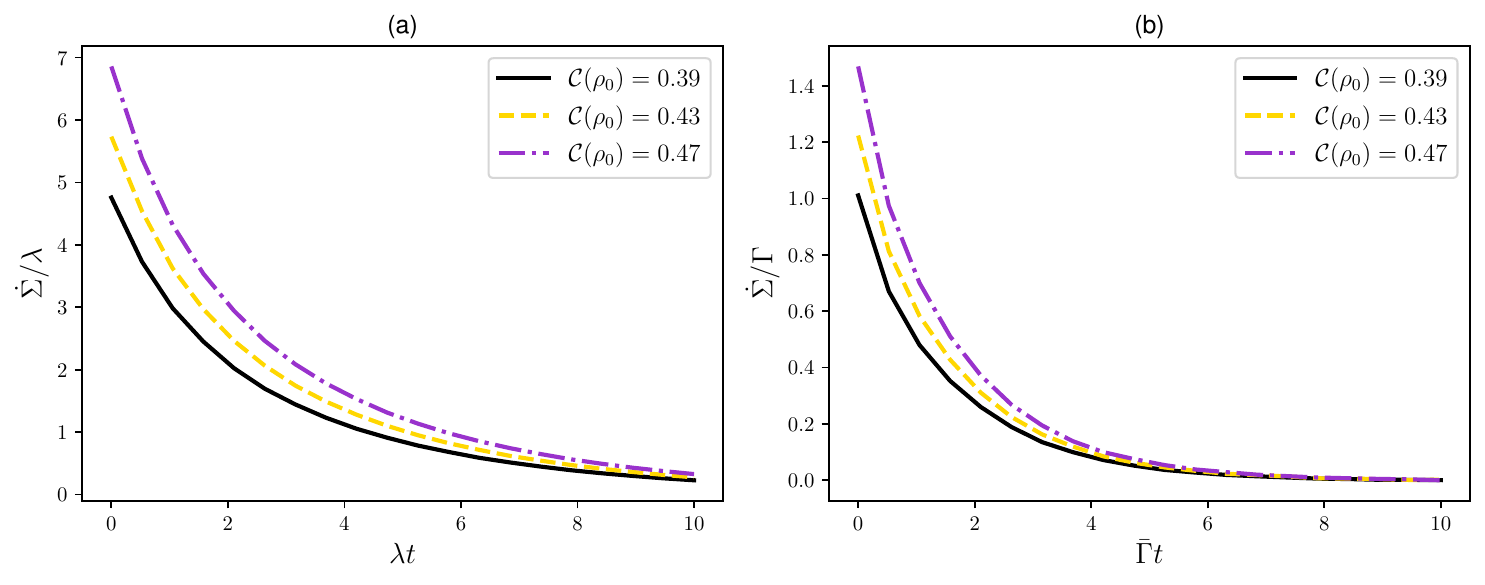}
\caption{Wehrl entropy production rate $\dot{\Sigma}$ as a function of time for a qutrit. In both Panels, we show several curves corresponding to different values of the initial coherence, as quantified by the $l_1$-norm, i.e. $\mathcal{C}(\rho_0) = 2(|\alpha|+|\beta|+|\gamma|)$. In Panel (a), we consider a single qutrit subject to pure dephasing, while in Panel (b) we consider we the case of amplitude damping dynamics. The initial states are randomly generated so that we reproduce the value of coherence shown in the legend. The average number of excitations in the bath is given by $\bar{n}=0.5$, while $\bar{\Gamma}=\Gamma (2\bar{n}+1)$.}
\label{fig:qutrit_EP_vs_t}   
\end{figure}

Furthermore, we can consider the case of a qutrit, i.e. a spin$-1$ particle. In this case, for the sake of simplicity, we can assume that all the entries of the density matrix are real numbers, i.e. 
\begin{align}
\rho_0 = \begin{pmatrix}
\rho_{00} & \alpha & \beta \\
\alpha & \rho_{11} & \gamma \\
\beta & \gamma & \rho_{22} 
\end{pmatrix} \, ,
\end{align}
thus, according to the $l_1$-norm, the quantum coherence associated with it is given by $\mathcal{C}(\rho_0) = 2(|\alpha|+|\beta|+|\gamma|)$. In this case, we can represent the spin operators as
\begin{equation}
\label{eq:spin_operators}
J_x = \frac{1}{\sqrt{2}} \begin{pmatrix}
0 & 1 & 0 \\
1 & 0 & 1 \\
0 & 1 & 0
\end{pmatrix}, \qquad
J_y = \frac{1}{\sqrt{2}} \begin{pmatrix}
0 & - i & 0 \\
i & 0 & -i \\
0 & i & 0
\end{pmatrix}, \quad
J_z = \begin{pmatrix}
1 & 0 & 0 \\
0 & 0 & 0 \\
0 & 0 & -1
\end{pmatrix},
\end{equation}
which can be used to construct the dissipators of Eqs.~(\ref{eq:diss_deph})-(\ref{eq:ampl_damp_diss}).
We can take $\ket{J}$ as $(1 \quad 0 \quad 0)^{\rm T}$, then construct the corresponding spin coherent state through the definition given by Eq.~(\ref{eq:spin_coherent_state_def}), whence, by resorting to Eq.~(\ref{eq:HusimiQ}), we obtain the following expression for the Husimi Q-function
\begin{eqnarray}
\mathcal{Q}(\Omega) & = & \rho_{00} \cos^4{\left ( \frac{\theta}{2} \right )} + \rho_{11} \left (\frac{\sin^2{\theta}}{2} \right) +
\rho_{22} \sin^4{\left ( \frac{\theta}{2} \right )} \nonumber \\
&\times& \frac{\sqrt{2}}{2}\sin{\theta} \left [ a+c +(a-c)\cos{\theta} \right] \cos{\phi} + \frac{b}{2} \sin^2{\theta} \cos{(2 \phi)} .
\end{eqnarray}
This function, together with its derivatives, can be used to calculate the entropy production rate for the cases of dephasing and amplitude damping channels, through Eqs.~(\ref{eq:EP_deph})-(\ref{eq:EP_ampl_damp}). Similarly to the case of a single qubit, one can track the entropy production rate over time, i.e. $\dot{\Sigma} = \dot{\Sigma}(t)$. One needs to explicitly solve the dynamics and calculate the corresponding entropy production rate at each time step by using Eqs.~(\ref{eq:EP_deph})-(\ref{eq:EP_ampl_damp}), and integrating over the phase space. For both the dephasing and the amplitude damping channels, one can prepare the system into different initial states (characterised by a different amount of coherence) and determine $\dot{\Sigma} = \dot{\Sigma}(t)$. In Fig.~\ref{fig:qutrit_EP_vs_t} we show that initial states characterised by a larger coherence correspond to higher entropy production rates.

\section{Conclusions}

We discussed how entropy production can be used to characterise and quantify the irreversibility  arising when we consider a quantum system weakly interacting with a thermal bath. More specifically, beside the standard approach based on von Neumann entropy, we resorted to a phase-space formulation of  the entropy production. Such description proved useful to attack the case of spin systems that can be described by means of the the so-called spin coherent states. We proved that -- both for dephasing and amplitude damping channels -- the initial coherence available to the system is monotonically related to the entropy production rate. However, such a relationship is not general. For instance, as shown in Ref.~\cite{Zicari:2023}, for randomly generated initial states of bipartite spin systems, there might be instances contradicting the direct proportionality between initial coherence and high entropy production rates.

\section*{Acknowledgments}
B\c{C} is partially supported by The Scientific and Technological Research Council of Turkey (TUBITAK) under Grant No.~121F246.
MP acknowledges the support by the European Union's Horizon 2020 FET-Open project  TEQ (766900), the Horizon Europe EIC Pathfinder project QuCoM (Grant Agreement No.~101046973), the Leverhulme Trust Research Project Grant UltraQuTe (grant RGP-2018-266), the Royal Society Wolfson Fellowship (RSWF/R3/183013), the UK EPSRC (EP/T028424/1), and the Department for the Economy Northern Ireland under the US-Ireland R\&D Partnership Programme. 


\end{document}